\documentclass[11pt,review,sort&compress]{elsarticle}

\usepackage{latexsym}
\usepackage{amssymb}
\usepackage{epsfig,amsmath,graphics}
\usepackage{epstopdf}
\usepackage{verbatim}
\usepackage[hypertex]{hyperref}
\usepackage{graphicx}
\usepackage{subfigure}

\begin{document}


\title{Semiconductor Probes of Light Dark Matter}

\author[stan]{Peter W. Graham}
\ead{pwgraham@stanford.edu}
\author[jhu]{David E. Kaplan}
\ead{dkaplan@pha.jhu.edu}
\author[stan,jhu]{Surjeet Rajendran}
\ead{surjeet@stanford.edu}
\author[jhu]{Matthew T. Walters}
\ead{mwalters@pha.jhu.edu}

\address[stan]{Stanford Institute for Theoretical Physics, Department of Physics, Stanford University, Stanford, CA 94305-4060, USA}
\address[jhu]{Department of Physics and Astronomy, Johns Hopkins University, Baltimore, MD 21218-2608, USA}

\begin{abstract}

Dark matter with mass below about a GeV is essentially unobservable in conventional direct detection experiments. However, newly proposed technology will allow the detection of single electron events in semiconductor materials with significantly lowered thresholds. This would allow detection of dark matter as light as an MeV in mass. Compared to other detection technologies, semiconductors allow enhanced sensitivity because of their low ionization energy around an eV. Such detectors would be particularly sensitive to dark matter with electric and magnetic dipole moments, with a reach many orders of magnitude beyond current bounds. Observable dipole moment interactions can be generated by new particles with masses as great as $\sim 10^3$ TeV, providing a window to scales beyond the reach of current colliders.

\end{abstract}

\maketitle

\section{Introduction}
\label{Sec:Intro}

The particle nature of dark matter (DM) has been well established by astronomical and cosmological data (\cite{BertoneEtAl,Bergstrom} and references therein).  It is reasonable to expect the DM particle to carry non-gravitational interactions. These non-gravitational interactions may permit direct detection of DM,  plausibly leading to a deeper understanding of its origins and the structure of particle physics. A good case can be made for weak-scale interactions between the standard model and DM. The possible existence of new states at the weak scale, as suggested by the hierarchy problem, could lead to such interactions. A variety of experiments are currently probing these interactions.  These experiments measure the energy deposited by DM as it scatters off the atoms in a detector.  Current experiments are sensitive to recoil energies $\gtrsim$ keV \cite{Gaitskell}. At these recoil energies, the experiments are dominantly sensitive to the scattering of DM particles with masses larger than $\sim 1$ GeV off the atomic nucleus \cite{BanksEtAl}. 

A lighter particle bound to the DM halo is kinematically forbidden from depositing energies greater than a keV. Owing to our ignorance of the physics responsible for DM, it is desirable to develop technological tools to explore all possible regions of the DM parameter space. The ability to detect low energy ($\sim$ eV) electron recoil events will significantly extend our reach into this parameter space, as was demonstrated recently in \cite{EssigEtAl}. The energy deposited by a light DM particle on the nucleus is suppressed by the nuclear mass. The difficulty of detecting such low energy nuclear recoils is further complicated by the anaemic response of the nucleus to such events. However, since the electron is light, DM can dump more energy into it. Further, energy deposition into electrons can lead to more readily identifiable events such as ionization in the detector. Technological advances in semiconductor-based DM detectors have made the detection of such ionization events a realistic possibility. Strategies to suppress backgrounds to ultra low levels similar to typical direct detection experiments also seem feasible \cite{PyleEtAl,Wang}. One possible experiment of this type is the CDMSLite proposal, which would modify existing CDMS technology to reduce the energy threshold by approximately three orders of magnitude, thereby allowing detection of single electron recoils \cite{CabreraEtAl}.  

The ionization of electrons always involves transfer of momentum from DM to the atomic nucleus. The cross-section for such interactions is suppressed by a form factor for momentum transfers much bigger than the inverse Bohr radius of the concerned electron. The electrons in semiconductors and noble gas detectors have comparable Bohr radii. However, the bandstructure of the semiconductor allows for the ionization of electrons with relatively lower energy ($\sim 1$ eV) in comparison with the energy needed to ionize electrons in a noble gas detector ($\sim 10$ eV). Since the DM particle has to lose more energy in the case of a noble gas detector, the momentum transferred to the nucleus is higher, leading to a form factor suppression of the cross-section. Owing to the smaller energy transferred in the case of the semiconductor, the momentum transferred to the nucleus is also smaller, leading to an unsuppressed cross-section. 

In this paper, we argue that semiconductor detectors able to measure the production of single electron-hole pairs have the potential to detect light DM in a wide range of parameter space, orders of magnitude beyond current bounds. We also show that such semiconductor devices possess an enhanced sensitivity to light DM in comparison with noble gas detectors. We begin, in section \ref{Sec:Models}, by considering the possible models for DM-electron interactions, either through renormalizable couplings or effective operators such as electromagnetic dipoles. We illustrate, through the aid of a simple concrete example, the ease with which these electromagnetic moments are generated for DM particles as a consequence of new states at the weak scale. In section \ref{Sec:Rate}, we compute the rate for a light DM particle to scatter off a valence electron bound to a semiconductor. In computing this rate for various operators, we focus in particular on electromagnetic moments because they are both easily generated in many models and have enhanced cross-sections due to their coupling to a long range force carrier, namely, the photon. Consequently, these operators may offer the easiest way to probe the existence of such light DM. Using an estimate of the possible backgrounds at CDMSLite, we examine the sensitivity of such a device to DM-electron interactions in section \ref{Sec:Sensitivities}, in comparison to current bounds on these operators, as well as the limits possible with noble gas detectors.

\section{Models}
\label{Sec:Models}

In order to study the potential reach of detectors such as CDMSLite, we must consider the possible operators that generate DM-electron interactions, as well as the current constraints on the corresponding parameters. We restrict ourselves to the simplest extensions to the standard model, but the DM sector could contain other scenarios (such as \cite{Khlopov,KhlopovEtAl,KrnjaicEtAl,BoehmEtAl,BoehmFayet}), possible signals of which should be studied in future work. We also focus specifically on the case of light DM ($m_e \lesssim m_\chi \lesssim$ 10 GeV), in order to find unexplored parameter space for these simple models.

\subsection{Dipole Moments}

The simplest extension to the standard model is for DM and electrons to interact electromagnetically. Due to constraints on the possible electric charge of DM \cite{McDermottEtAl}, the lowest dimensionality available for electromagnetic interactions corresponds to the dimension-five dipole moment operators

\begin{equation}
\mathcal{L}_{dipole} = -\frac{i}{2} \bar{\chi} \sigma^{\mu \nu}(\mu_\chi + d_\chi \gamma^5) \chi F_{\mu \nu} \textrm{,}
\end{equation}

\begin{figure}[t]
\centering
\includegraphics[width=10cm]{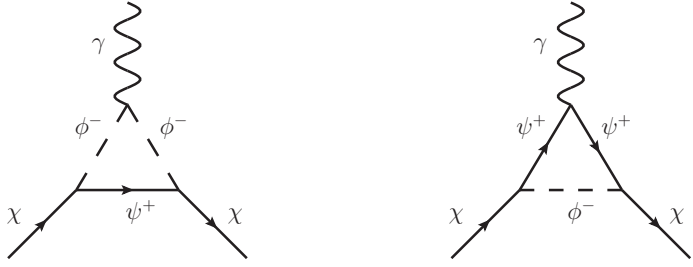}
\caption{One-loop contributions to DM dipole moment due to a charged fermion-scalar pair.}
\label{fig:dipoles}
\end{figure}

{\noindent}where $\mu_\chi$ and $d_\chi$ correspond to the DM magnetic and electric dipole moments, respectively. These dipole moments correspond to a cutoff scale ($d_\chi \sim \Lambda^{-1}$) and arise from loop interactions involving heavy charged particles, such as those shown in Figure \ref{fig:dipoles}. These dipole moments are then easily generated in models where DM carries a conserved charge, such as asymmetric dark matter \cite{Nussinov,Kaplan,KitanoEtAl,KaplanEtAl,FalkowskiEtAl}. As an illustrative example, we consider two new heavy intermediaries (a fermion and a scalar), in the limit where both have approximately the same mass $M$ and coupling to DM $g$. We then obtain (based on calculations similar to \cite{CheungEtAl}) the dipole moment

\begin{equation}
d_\chi \approx \frac{e g^2}{8 \pi^2 M} \textrm{,}
\end{equation}

\begin{figure}[t]
\centering
\includegraphics[width=6cm]{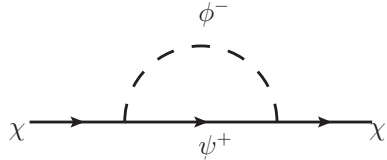}
\caption{One-loop contribution to DM mass due to a charged fermion-scalar pair.}
\label{fig:massloop}
\end{figure}

{\noindent}with the same approximate form for $\mu_\chi$. There might be some worry that loops involving these heavy charged intermediaries, shown in Figure \ref{fig:massloop}, would push the natural DM mass beyond the MeV or GeV scale. The contribution from this diagram is

\begin{equation}
\delta m_\chi \approx \frac{g^2 M}{16 \pi^2} \textrm{.}
\end{equation}

The important feature of this expression is that decreasing the coupling $g$ between DM and the heavy intermediaries decreases the effective scale contributing to the DM mass. However, this decrease in $g$ actually increases the effective scale contributing to the DM dipole moment. This means that for a generic set of heavy charged intermediaries, a large effective dipole scale does not imply a large mass contribution, provided the coupling with DM is small. For example, a charged fermion-scalar pair with $M \approx 500$ GeV and $g \approx 0.2$ would contribute $\delta m_\chi \approx 100$ MeV and $d_\chi \approx 3 \times 10^{-4}$ TeV$^{-1}$. As we will show in section \ref{Sec:Sensitivities}, the enhanced cross-sections of dipole interactions at low momentum transfer make them the strongest candidate for detection with CDMSLite, with experimental sensitivity to effective mass scales $\lesssim$ 10$^3$ TeV.

\subsection{Effective Pointlike Vertex}

The next simplest extension is the dimension-six effective four-fermion vertex, which corresponds to the exchange of a very massive mediator (such as a scalar or vector) which is then integrated out of the theory. An example is the vector-channel operator

\begin{equation}
\mathcal{L}_{point} = \frac{1}{\Lambda^2} \bar{\chi} \gamma^\mu \chi \bar{\psi}_e \gamma_\mu \psi_e \textrm{,}
\end{equation}

{\noindent}where $\Lambda$ corresponds to the cutoff of this effective theory (roughly the mass of the intermediary particle). The strongest constraints on these pointlike interactions come from collider experiments such as LEP \cite{FoxEtAl,GoodmanEtAl,MambriniEtAl}. For example, the vector interaction above is currently restricted to a cutoff scale $\Lambda \gtrsim 480$ GeV. Calculations based on the method presented in section \ref{Sec:Rate} indicate that CDMSLite would only be able to search for pointlike interactions up to the scale $\Lambda \approx 200$ GeV. The weakness of this projected sensitivity is due to the lack of enhancement for DM pointlike scattering at low recoil energies. Semiconductor detectors will then be no more sensitive to dimension-six DM interactions than collider experiments. However, higher scale physics which generates dimension-six operators will very often generate dipole moment operators, as well, whose recoils suffer from less suppression. As will be shown in section \ref{Sec:Sensitivities}, dipole operators will therefore allow CDMSLite to probe scales far beyond the reach possible with pointlike interactions, either in direct detection or collider experiments. In light of these facts, we do not consider pointlike interactions for the remainder of this paper.

\subsection{Broken $U(1)$}

The final possibility for simply extending the standard model is to introduce new light particles to the theory. In this case, DM can interact via a broken $U(1)$ gauge interaction, with the corresponding dimension-four operator

\begin{eqnarray}
\mathcal{L}_{A'\chi} = g_\chi A'_\mu \bar{\chi} \gamma^\mu \chi \textrm{,}
\end{eqnarray}

{\noindent}where $A'_\mu$ is the DM gauge field. A DM-electron interaction could then result from kinetic mixing of amplitude $\epsilon$ between $A'_\mu$ and the standard model photon, which can be diagonalized to give a DM gauge boson-electron coupling term

\begin{eqnarray}
\mathcal{L}_{A'e} = -\epsilon e A'_\mu \bar{\psi}_e \gamma^\mu \psi_e \textrm{.}
\end{eqnarray}

This interaction also has a large parameter space available for semiconductor detectors, which was discussed in \cite{EssigEtAl}. We consider this interaction later in some detail for the sake of of thoroughness and comparison with that work, but our main focus is on the potential exploration of new physics through DM dipole moments.

\section{Detection Rate}
\label{Sec:Rate}

As a simpler conceptual example, we first consider DM ionizing a single isolated atom. We then turn to DM interacting with a semiconducting lattice to excite an electron from the valence band to the conduction band. Our approaches to these two cases are rather similar, as are the resulting cross-sections. This topic is similar to \cite{EssigEtAl,KoppEtAl,DedesEtAl}, but differs in the treatment of the lattice bandstructure and in the inclusion of momentum transfer to nuclei.

\subsection{Basic Kinematics}

First, we briefly review the kinematics of electron recoil interactions. Our convention for momenta is to use $\vec{p}$ to indicate incoming momenta and $\vec{k}$ for outgoing momenta. Also, since the DM velocity $v_\chi \sim 10^{-3}$, we use simpler nonrelativistic kinematics.

Interactions are classified by the recoil energy $E_R$, defined as the kinetic energy of the outgoing electron

\begin{equation}
E_R = \frac{k_e^2}{2 m_e} \textrm{.}
\end{equation}

In the lab frame, the electron (initially in a bound state) can have a nonzero incoming momentum $\vec{p}_e$, with the probability for this momentum determined by the electron's momentum space wavefunction $\tilde{\psi}$. However, the initial energy for the electron is simply $-E_B$, the binding energy associated with its initial state $\psi$. We also choose the lab frame to be such that the nucleus has no initial momentum ($\vec{p}_N = 0$).

The momenta values we consider are much lower ($\sim$ keV) than the nuclear masses of silicon and germanium, such that the final kinetic energy of the nucleus can be neglected. The resulting energy conservation equation can be rewritten as

\begin{equation}
k_\chi^2 = p_\chi^2 - 2 m_\chi (E_R + E_B) \textrm{.}
\end{equation}

We can then calculate the minimum possible momentum transfer $q$ necessary to ionize an electron with binding energy $E_B$. For the case of semiconductors, with $E_B \sim 1$ eV, the DM must at least transfer momentum $q \sim 1$ keV for ionization to be possible. For noble gases, with a larger $E_B \sim 10$ eV, the mininum transfer necessary is $q \sim 10$ keV. This increase in momentum transfer away from the inverse Bohr radius and into the form factor regime suppresses the noble gas detection rates, reducing their experimental sensitivity in comparison with semiconductors. This suppression in the case of dipole moment interactions is given in detail in equations (\ref{eqn:EDMform}) and (\ref{eqn:MDMform}).

\subsection{Single Atom Ionization}

Our example initial state consists of free DM and a bound hydrogenic atom. If the atom is ionized by the DM-electron interaction, then the final state consists of the recoiling DM, escaping electron, and remaining nucleus. For calculational simplicity, we model the nuclear final state as a plane wave. However, the final state of the nucleus is in fact not relevant to the cross-section. Rather, what matters is that the nucleus, due to its large mass, can absorb momentum at negligible energy cost. From the perspective of the DM-electron system, the recoiling nucleus then breaks momentum conservation while preserving energy conservation.

The electron's resulting wavefunction is also deformed by the presence of the charged nucleus, causing it to deviate from a simple plane wave. This deformation causes a substantial enhancement to the cross-section, which can be approximated by combining a plane wave final state with a momentum-dependent enhancement factor. This factor is similar to the standard treatments of beta decay \cite{Morita} and Sommerfeld enhancement \cite{Sommerfeld} (for a clear review, see \cite{ArkaniEtAl}), as well as the work on noble gas detectors in \cite{EssigEtAl}. This enhancement factor can be found by exactly solving the Dirac equation for a free electron in the presence of a Coulomb potential, and comparing the solution to that of a plane wave, yielding

\begin{equation}
F(k_e) = \frac{2 \pi \nu}{1-e^{-2 \pi \nu}} \textrm{,}
\end{equation}

{\noindent}where $\nu$ is the $k_e$-dependent factor

\begin{equation}
\nu = \frac{Z_{eff} m_e \alpha}{k_e} \textrm{.}
\end{equation}

In order to calculate the cross-section for DM to ionize the atom, we then simply need to look at the usual scattering formula \cite{Sakurai}

\begin{equation}
d\sigma = \frac{1}{|v_{rel}|} \left( \displaystyle\prod_{f} \frac{d^3k_f}{(2 \pi)^3} \right) 2 \pi \, \delta(E_f - E_i) \left| \langle f|H|i \rangle \right|^2 \textrm{.}
\end{equation}

Using this formula, we can then write out the full cross-section for our ionization process, assuming (as stated earlier) that the lab frame corresponds to the nucleus rest frame and using the approximation $m_e << m_N$,

\begin{equation}
d\sigma = \frac{F(k_e)}{|\vec{v}_\chi|} \: \frac{d^3k_\chi}{(2\pi)^3} \: \frac{d^3k_e}{(2\pi)^3} \: \frac{d^3k_N}{(2\pi)^3} (2\pi)^4 \delta^4(k_f - p_i) \left| \tilde{\psi}(\vec{p}_\chi - \vec{k}_\chi - \vec{k}_e) \right|^2 \left| \tilde{H}_{int}(\vec{p}_\chi - \vec{k}_\chi) \right|^2 \textrm{,}
\end{equation}

{\noindent}where $\tilde{\psi}$ and $\tilde{H}_{int}$ refer to the Fourier transforms of the initial electron bound state wavefunction and DM-electron interaction Hamiltonian, respectively. For our initial bound state, we use the hydrogen ground state wavefunction

\begin{equation}
\psi_{0}(\vec{x}) = \frac{1}{\sqrt{\pi a^3}} e^{-r/a} \textrm{,}
\end{equation}

{\noindent}where $a$ is the Bohr radius.

We now calculate the cross-sections for the interactions considered earlier, the first of which is the electric dipole moment (EDM) interaction. Averaging over initial spins, summing over final spins, and using the approximation $k_e << q$ (for details, see the appendix), we obtain the approximate cross-section

\begin{equation}
\left. \frac{d \sigma}{dE_R} \, \right|_{EDM} \, \approx \, \frac{16 a^2 d_{\chi}^2 k_e F(k_e)}{\pi v_{\chi}^2} \, \left[ \ln \left( \frac{1+a^2q_{min}^2}{a^2q_{min}^2} \right) - \frac{6a^4q_{min}^4 + 15a^2q_{min}^2 + 11}{6(1+a^2q_{min}^2)^3} \right] \textrm{.}
\end{equation}

The term $v_{\chi}$ refers to the incoming velocity of the DM particle (in the lab frame), and the various momenta have the following definitions

\begin{equation}
\begin{split}
k_e &= \sqrt{2 m_e E_R} \textrm{,} \\
q_{min} &= m_\chi v_\chi - \sqrt{m_\chi^2 v_\chi^2 - 2 m_\chi (E_R + E_B)} \textrm{.}
\end{split}
\end{equation}

In order to understand the form factor suppression for momentum transfer above $a^{-1}$, we can take the further limit of $aq_{min}>>1$, obtaining

\begin{equation}
\left. \frac{d \sigma}{dE_R} \, \right|_{EDM} \longrightarrow \, \frac{4 d_{\chi}^2 k_e F(k_e)}{\pi v_{\chi}^2 a^6 q_{min}^8} \textrm{.}
\label{eqn:EDMform}
\end{equation}

The second cross-section corresponds to the similar magnetic dipole moment (MDM) interaction. Using the same approximations and variables as the EDM case, we find the cross-section

\begin{equation}
\left. \frac{d \sigma}{dE_R} \, \right|_{MDM} \, \approx \, \frac{64 \alpha^2 a^2 \mu_\chi^2 k_e F(k_e)}{3 \pi v_\chi^2} \, \left[ \ln \left( \frac{1+a^2q_{min}^2}{a^2q_{min}^2} \right) - \frac{12a^4q_{min}^4 + 30a^2q_{min}^2 + 19}{12(1+a^2q_{min}^2)^3} \right] \textrm{.}
\end{equation}

Not surprisingly, this cross-section is quite similar to that of an EDM, but is suppressed by an additional approximate factor of $\alpha^2$. This factor corresponds to the average velocity of the bound electron ($\langle v_e \rangle \sim \alpha$). We can also obtain the similar form factor

\begin{equation}
\left. \frac{d \sigma}{dE_R} \, \right|_{MDM} \longrightarrow \, \frac{16 \alpha^2 \mu_\chi^2 k_e F(k_e)}{3 \pi v_\chi^2 a^4 q_{min}^6} \textrm{.}
\label{eqn:MDMform}
\end{equation}

The final cross-section corresponds to the broken $U(1)$ DM gauge interaction, for which we consider two limiting regimes. The first corresponds to a heavy mediator, with mass $m_A$ much greater than the momentum transfer, such that the interaction is effectively pointlike. In this limit, the cross-section is

\begin{equation}
\left. \frac{d \sigma}{dE_R} \, \right|_{heavy} \, \approx \, \frac{128 \lambda^2 k_e F(k_e)}{3 v_{\chi}^2 m_A^4} \, \frac{1}{(1+a^2q_{min}^2)^3} \textrm{,}
\end{equation}

{\noindent}where $\lambda$ is the effective DM-electron coupling

\begin{equation}
\lambda=\epsilon \sqrt{\frac{g_\chi^2}{4 \pi}} \textrm{.}
\label{eqn:lambda}
\end{equation}

The second limit corresponds to a light mediator, with mass much less than the momentum transfer, such that $A'_\mu$ is effectively massless. This yields the cross-section

\begin{equation}
\left. \frac{d \sigma}{dE_R} \right|_{light} \approx \frac{128 \lambda^2 a^4 k_e F(k_e)}{3 v_{\chi}^2} \left[ \frac{3}{a^2q_{min}^2} + \frac{9(1+a^2q_{min}^2)^2 + 3(1+a^2q_{min}^2) + 1}{(1+a^2q_{min}^2)^3} - 12 \ln \left( \frac{1+a^2q_{min}^2}{a^2q_{min}^2} \right) \right] \textrm{.}
\end{equation}

\subsection{Semiconductor Valence Band}

We now consider an atom in a semiconducting lattice. For an electron in any periodic potential, the delocalized eigenfunctions of the Hamiltonian can be expressed in terms of localized wavefunctions, in the form \cite{AshcroftEtAl,BohmEtAl}

\begin{equation}
\psi_{\vec{b}}(\vec{x}) = \frac{1}{\sqrt{N}} \displaystyle\sum_{n} e^{i\vec{b} \cdot \vec{x}_n} \phi(\vec{x}-\vec{x}_n) \textrm{,}
\end{equation}

{\noindent}where $N$ is the total number of lattice sites and $\vec{b}$ is a wavevector with components related to the dimensions $L_i$ of the lattice by the relationship

\begin{equation}
b_i = \frac{2 \pi n_i}{L_i} \textrm{,}
\end{equation}

{\noindent}where $i$ runs over the values $x$, $y$, and $z$, and $n_i$ is any integer from 1 to $N_i$, the number of lattice sites in the $i$-direction ($N=N_x N_y N_z$). The wavefunction $\phi(\vec{x}-\vec{x}_n)$ is a localized Wannier wavefunction centered around each individual lattice site (located at $\vec{x}_n$).

In the tight-binding, linear combination of atomic orbitals (LCAO) approximation \cite{YuEtAl}, the Wannier wavefunctions $\phi$ are written in the basis of free atomic orbitals. These coefficients are very small for all atomic orbitals except those near the bound-state energy of $\psi_{\vec{b}}$ \cite{AshcroftEtAl}, which for valence band states are the highest occupied $s$- and $p$-states. These outermost states can be reasonably approximated with hydrogenic wavefunctions, due to the screening of the nuclear charge by the inner core electrons.

For the final state, the valence band electron is excited into the conduction band, where it can be treated as an approximately free electron, with two corrections. Due to the weak periodic potential of the lattice, a conduction band electron propagates with the effective mass $m_e^* = f_e m_e$. The correction factor $f_e$ is an element-dependent factor determined by the direction of the electron momentum in the lattice and the energy curvature along the conduction band. We use the density of states average values for $f_e$ corresponding to the very edge of the conduction band. This gives an approximate estimate for the interaction cross-section, which will only be slightly modified by a more exact calculation. We use $f_e = 1.1$ for silicon and $f_e = 0.6$ for germanium \cite{Boer,Kittel}.

The second correction comes from the presence of the positively charged hole remaining in the valence band. The Coulomb interaction between these charges causes the same enhancement as the atomic case, where the effective charge $Z_{eff}$ felt by the outgoing electron is simply that of the remaining hole. 

With these particular initial and final states, we can use a similar approach to the free atom considered earlier, writing the $\vec{b}$-dependent interaction cross-section

\begin{equation}
d\sigma_{\vec{b}} = \frac{F(k_e)}{|\vec{v}_\chi|} \: \left( \displaystyle\prod_{f} \frac{d^3k_f}{(2 \pi)^3} \right) (2\pi)^4 \delta^4(k_f - p_i) \left| \tilde{\psi}_{\vec{b}}(\vec{p}_\chi - \vec{k}_\chi - \vec{k}_e) \right|^2 \left| \tilde{H}_{int}(\vec{k}_\chi - \vec{p}_\chi) \right|^2 \textrm{.}
\end{equation}

For interactions localized to a single lattice site (momentum transfer of $\mathcal{O}(a^{-1}$)), the cross terms for wavefunctions of different sites are negligible. This cross-section can then be expressed in terms of interactions with a single local Wannier wavefunction. We therefore approximate cross-sections involving highly delocalized states spread throughout the entire lattice by calculating cross-sections involving a single localized state, repeated periodically at the $N$ sites of the lattice,

\begin{equation}
d\sigma_{\vec{b}} \approx \frac{F(k_e)}{|\vec{v}_\chi|} \: \left( \displaystyle\prod_{f} \frac{d^3k_f}{(2 \pi)^3} \right) (2\pi)^4 \delta^4(k_f - p_i) \left| \tilde{\phi}(\vec{p}_\chi - \vec{k}_\chi - \vec{k}_e) \right|^2 \left| \tilde{H}_{int}(\vec{k}_\chi - \vec{p}_\chi) \right|^2 \textrm{.}
\end{equation}

Our full cross-section for a single lattice site is then simply the average of the individual $d\sigma_{\vec{b}}$,

\begin{equation}
d\sigma = \frac{1}{N} \displaystyle\sum_{\vec{b}} d\sigma_{\vec{b}} \textrm{.}
\end{equation}

In this approximation, the $\vec{b}$-dependence for each $d\sigma_{\vec{b}}$ is contained entirely in the initial state binding energy $E_B$. Our total cross-section then changes from an average over all possible $\vec{b}$ to an integral over all possible $E_B$,

\begin{equation}
d\sigma \approx \int dE_B \: \rho(E_B) d\sigma(E_B) \textrm{,}
\end{equation}

\begin{figure}[t]
\centering
\includegraphics[width=0.6\textwidth]{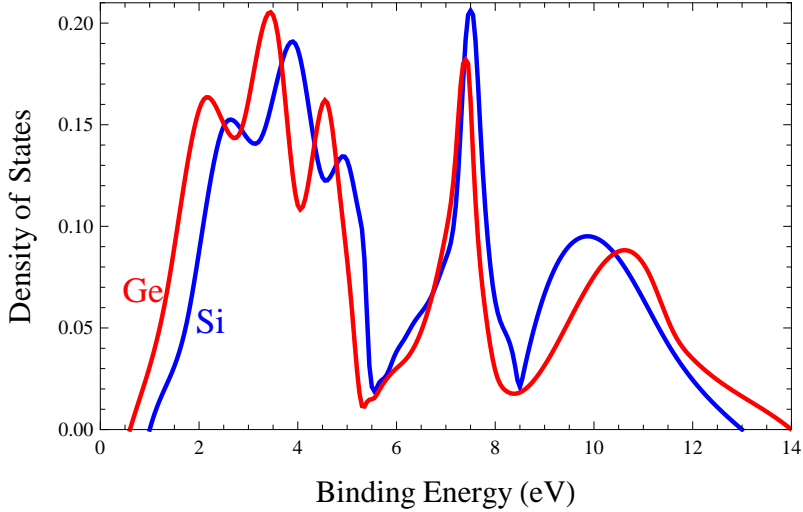}
\caption{Density of states as a function of binding energy for the valence bands of germanium (red) and silicon (blue), normalized such that $\int dE_B \, \rho(E_B)=1$.}
\label{fig:density}
\end{figure}

{\noindent}where $\rho(E_B)$ is an experimentally determined density of states (based on \cite{YuEtAl,ChadiEtAl} and shown in Figure \ref{fig:density}) accounting for the fact that some $E_B$ values correspond to more $\psi_{\vec{b}}$ states than others. This density of states then serves as an efficiency factor for scattering at various binding energies. For example, there is zero detection efficiency in germanium at the minimum $E_B$ of 0.7 eV, but the efficiency rapidly increases for slightly larger $E_B$. Note that our final result does not contain any directional dependence resulting from the lattice structure, but rather gives the directionally-averaged behavior of the total cross-section.

The cross-sections for electrons in a semiconducting lattice are then accurately approximated by those of a free hydrogenic bound state, in a weighted average over initial-state binding energies, with an altered final-state electron mass.

\subsection{Detection Rate}

Once the interaction cross-section is known, the total rate of detection (typically written in units of events/day/kg/eV) can be calculated using the following expression

\begin{eqnarray}
\frac{dR}{dE_R} = \frac{\rho_\chi \, \eta_e}{m_\chi} \int d^3v_\chi f(\vec{v}_\chi) v_\chi \frac{d\sigma}{dE_R} \textrm{,}
\end{eqnarray}

{\noindent}where $\rho_\chi$ is the DM mass density and $\eta_e$ is the valence electron number density per unit mass of the detector. For the DM density we use the value $\rho_\chi = 0.3$ GeV/cm$^3$ \cite{Gaitskell}. The function $f(\vec{v}_\chi)$ is the DM velocity distribution in the lab frame, meaning that we need to account for the Earth's average velocity $\vec{v}_E$ through the galaxy. We use the conventional Maxwellian distribution, truncated at a maximum escape velocity of $v_{esc}$ (in the average rest frame of the galaxy),

\begin{eqnarray}
f(\vec{v}_\chi) = \frac{1}{k} \exp \left( \frac{-(\vec{v}_\chi+\vec{v}_E)^2}{v_0^2} \right) \textrm{,}
\end{eqnarray}

{\noindent}where k is a normalization factor chosen such that $\int d^3v_\chi \, f(\vec{v}_\chi) = 1$,

\begin{eqnarray}
k = (\pi v_0^2)^{3/2} \left[ \textrm{erf} \left( \frac{v_{esc}}{v_0} \right) - \frac{2}{\sqrt{\pi}} \frac{v_{esc}}{v_0} e^{-v_{esc}^2/v_0^2} \right] \textrm{.}
\end{eqnarray}

We use the following values for the velocity parameters: average Earth velocity $v_E=240$ km/s, average DM velocity $v_0=230$ km/s, and DM escape velocity $v_{esc}=600$ km/s \cite{LewinEtAl}. These values give us $k = 2.504 \times 10^{-9}$ (in units with $c=1$).

Performing the angular integrals (of which our cross-sections are independent), we obtain the following expression

\begin{eqnarray}
\frac{dR}{dE_R} = \frac{\rho_\chi \, \eta_e \, \pi v_0^2}{m_\chi k v_E} \left[ \int_{v_{min}}^{v_{esc}+v_E} dv_\chi \, v_\chi^2 e^{-(v_\chi-v_E)^2/v_0^2}  \frac{d \sigma}{dE_R} - \int_{v_{min}}^{v_{esc}-v_E} dv_\chi \, v_\chi^2 e^{-(v_\chi+v_E)^2/v_0^2}  \frac{d \sigma}{dE_R} \right] \textrm{,}
\end{eqnarray}

{\noindent}where $v_{min}$ is the minimum DM velocity necessary to induce an interaction of recoil energy $E_R$,

\begin{eqnarray}
v_{min} = \sqrt{\frac{2 (E_R+E_B)}{m_\chi}} \textrm{.}
\end{eqnarray}

\section{Sensitivities}
\label{Sec:Sensitivities}

\subsection{Approach}

Using these detection rates, we can estimate the possible sensitivity for a detector such as CDMSLite.  We assume a flat background rate of 1 event/day/kg/keV, which was provided as an experimental estimation of the expected background \cite{CabreraEtAl}. This background rate is due to residual radioactivity, which produces high energy gamma rays that can Compton scatter in the detector. To determine if the experimental reach is limited by the background, we also include the sensitivity possible using germanium with no background. Due to the small number of events, we find that reducing the background beyond our estimate does not have a substantial effect on the exclusion sensitivity.

The proposed detection method involves the measurement of single electron-hole pairs, assuming that the energy deposited in the initial recoiling electron will prompt the formation of secondary electron-hole pairs. This method of detection will limit the energy resolution to the average energy per electron-hole pair, which is approximately 3 eV \cite{Klein,CabreraEtAl}. We do not consider this process in detail, and instead focus on the initial deposition of energy into a single electron. The possible interactions we consider are all peaked very strongly at low $E_R \sim$ eV, falling off quickly with growing recoil energy. Because of this, we focus solely on recoils with the lowest energy ($E_R < 9$ eV) where the signal is most competitive with the background. For CDMSLite, we assume the detection setup of CDMS II \cite{AhmedEtAl}, with 4.4 kg of germanium and 1.1 kg of silicon. We also consider the sensitivity of each material separately for the sake of comparison, rather than combine data for a total exclusion limit. Using the approach of \cite{FeldmanEtAl}, we can find the 95\% confidence level sensitivity possible after an experimental run of one year.

For comparison, we also include the possible sensitivity of measuring electron recoils in a xenon-based detector. This approximate sensitivity is calculated using the ionization cross-sections derived earlier, combined with atomic data from \cite{BungeEtAl}. For simplification, we only consider interactions with the valence electrons in the $5s$- and $5p$-states, which dominate the overall cross-section, and use $Z_{eff}=1$ for the Sommerfeld enhancement. We assume a detector mass of 1 kg, a runtime of 1 year, and negligible background, in order to find the maximum possible reach of noble gas detectors.

\subsection{Results}

\begin{figure}[t]
\centering
\begin{tabular}{cc}
\includegraphics[width=0.52\textwidth]{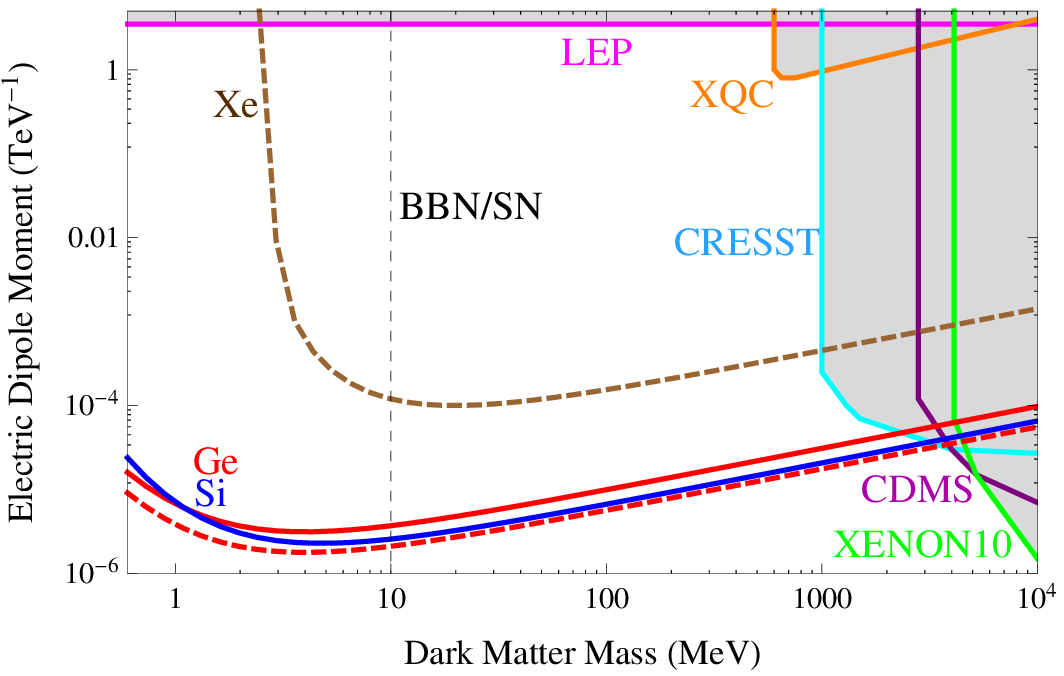}
\\
\includegraphics[width=0.52\textwidth]{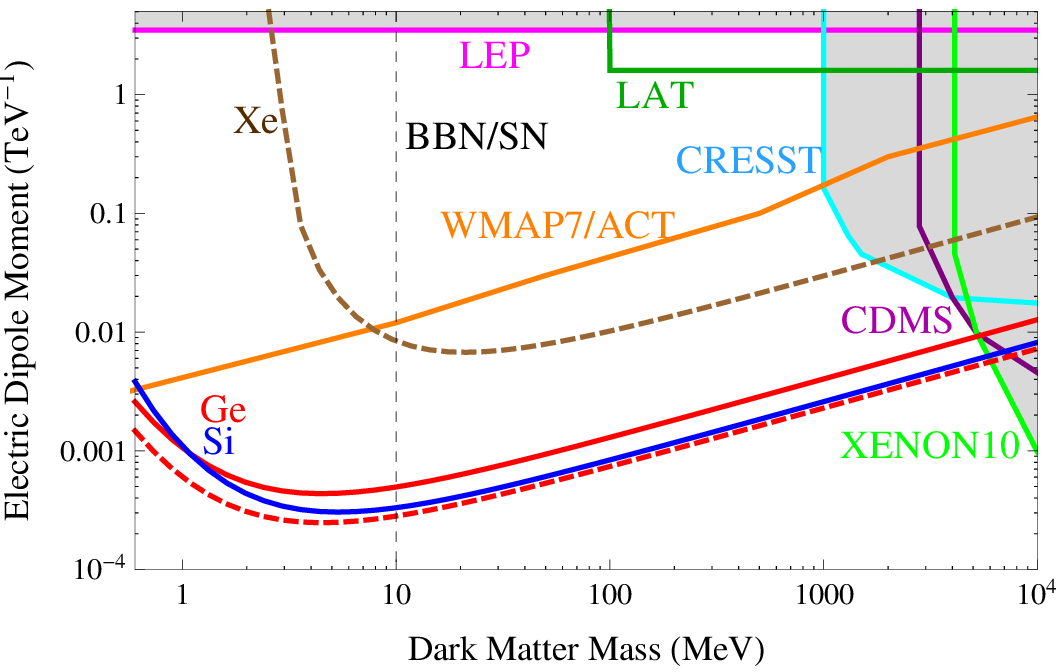}
\end{tabular}
\caption{Exclusion sensitivity at 95\% confidence level possible after 1 year, for (a) electric and (b) magnetic dipole moments. The solid lines assume a background of 1 event/day/kg/keV, while the dashed lines assume no background. Areas above the curves for germanium (red), silicon (blue), and xenon (brown) would be excluded. Regions in gray are already excluded for all models of DM by other experiments or astrophysical data. Masses to the left of the dashed black line are potentially constrained by supernova cooling and BBN. While a detailed calculation of these constraints on lighter masses is beyond the scope of this work, it is unlikely the entire region is fully excluded.}
\label{fig:dipolelimits}
\end{figure}

\begin{figure}[t]
\centering
\begin{tabular}{cc}
\includegraphics[width=0.5\textwidth]{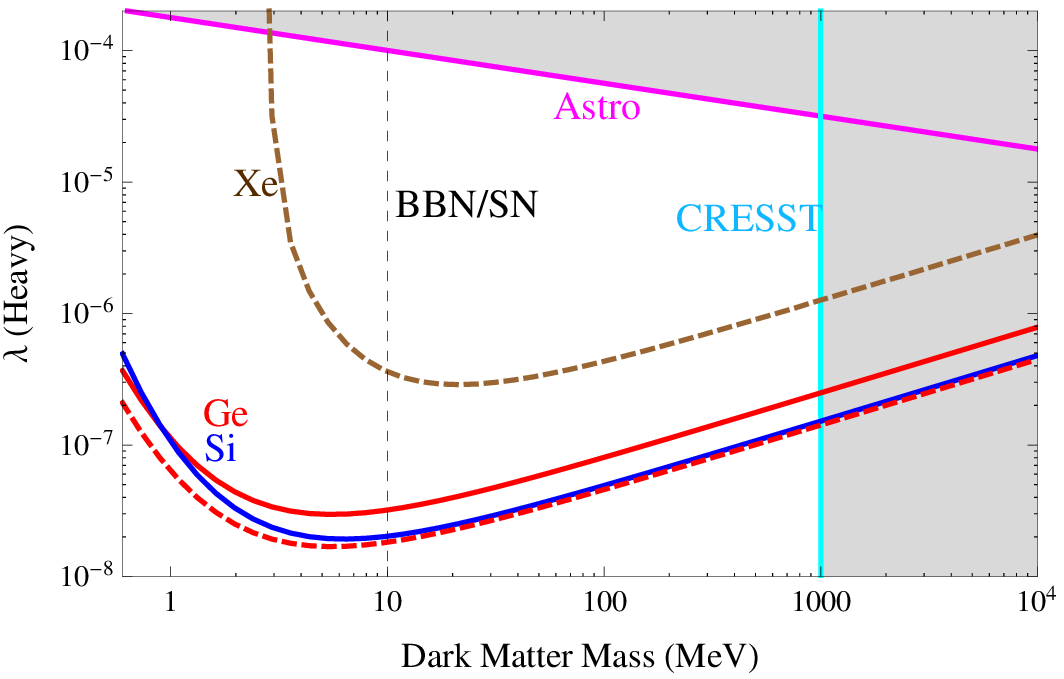}
\\
\includegraphics[width=0.5\textwidth]{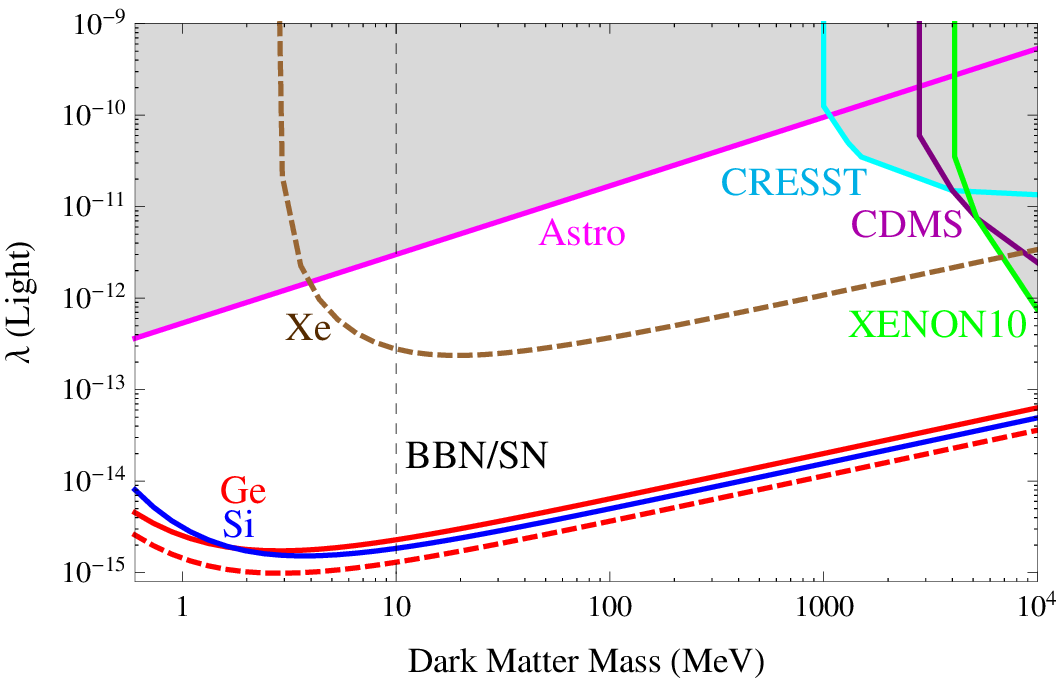}
\end{tabular}
\caption{Exclusion sensitivity at 95\% confidence level possible after 1 year, for effective $U(1)$ coupling $\left(\lambda = \epsilon \sqrt{\frac{g_\chi^2}{4\pi}} \right)$ with (a) $m_A = 10$ MeV and (b) $m_A = 1$ meV. The solid lines assume a background of 1 event/day/kg/keV, while the dashed lines assume no background. Areas above the curves for germanium (red), silicon (blue), and xenon (brown) would be excluded. Regions in gray are already excluded for all models of DM by other experiments or astrophysical data. Masses to the left of the dashed black line are potentially constrained by supernova cooling and BBN. While a detailed calculation of these constraints on lighter masses is beyond the scope of this work, it is unlikely the entire region is fully excluded.}
\label{fig:u1limits}
\end{figure}

\begin{figure}[t]
\centering
\begin{tabular}{cc}
\includegraphics[width=0.45\textwidth]{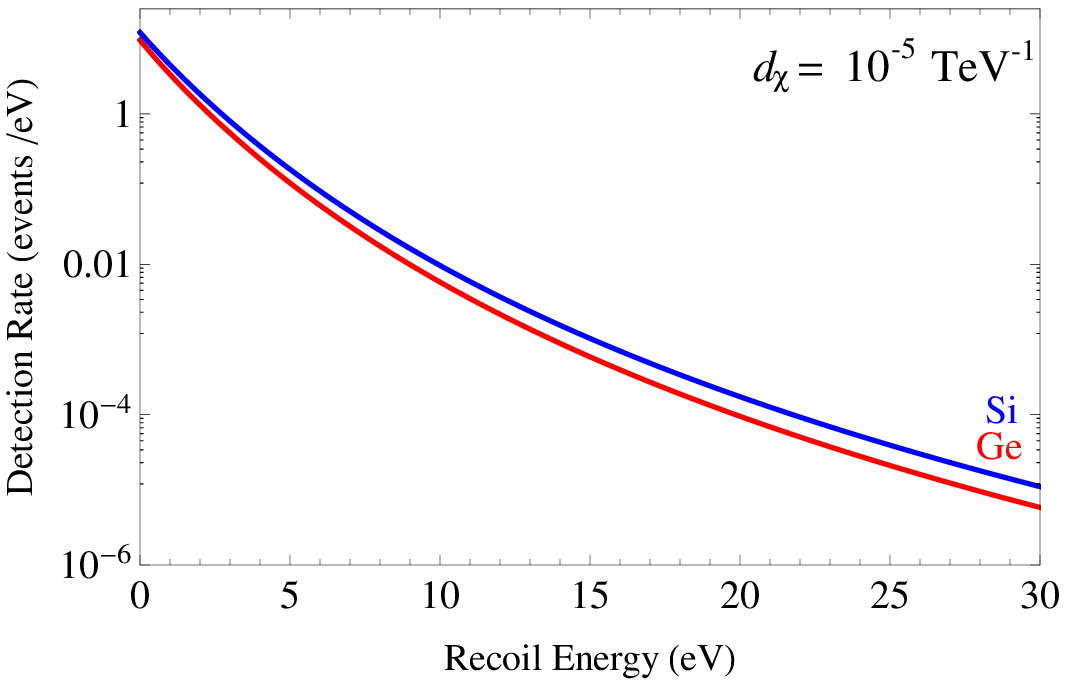}
&
\includegraphics[width=0.45\textwidth]{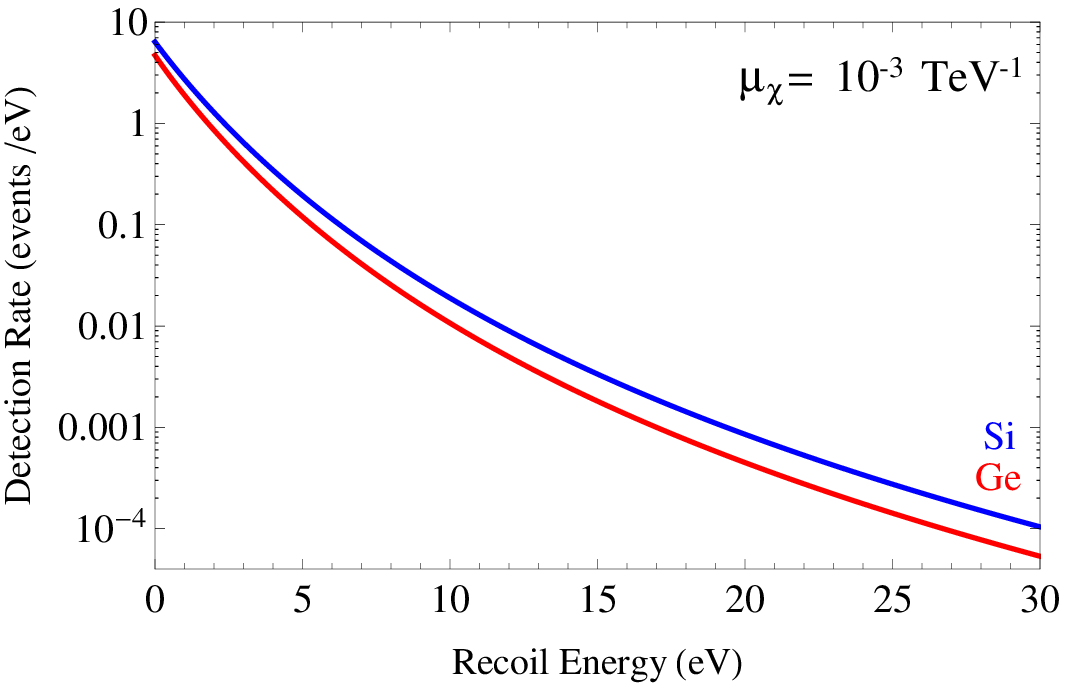}
\\
\includegraphics[width=0.45\textwidth]{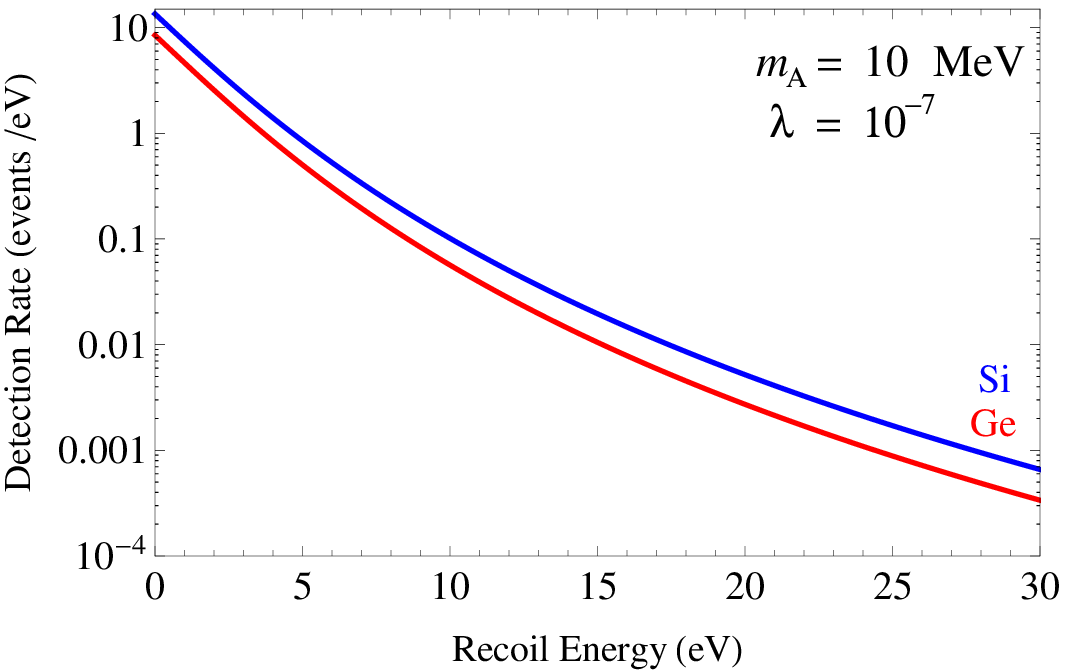}
&
\includegraphics[width=0.45\textwidth]{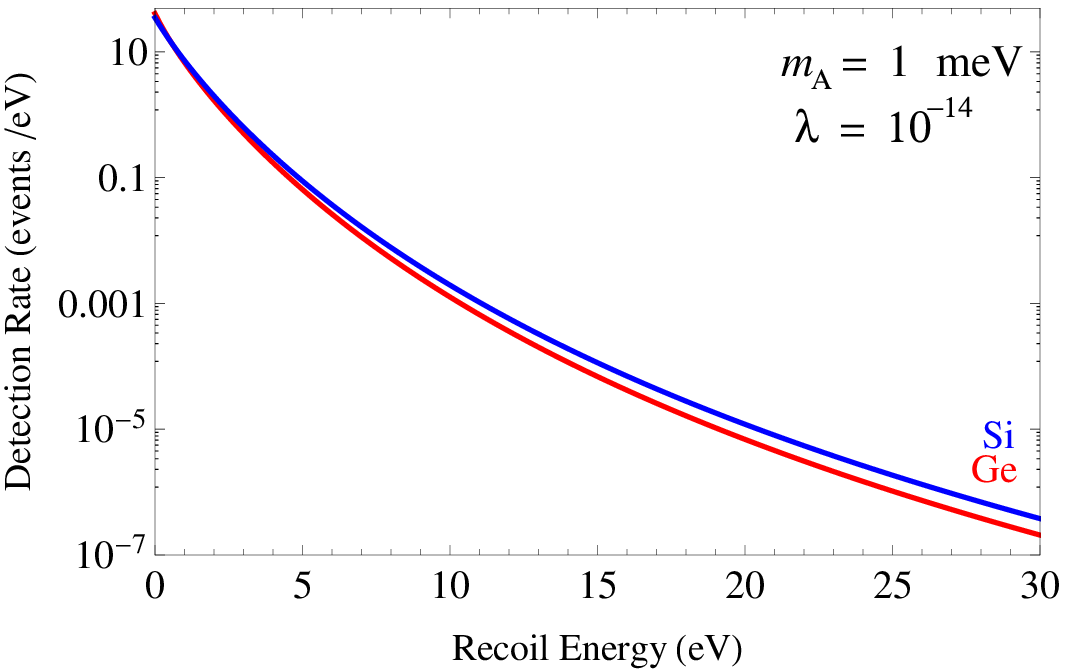}
\end{tabular}
\caption{Detection rates (in events/eV) after 1 year, for $m_\chi = 100$ MeV. These rates assume detector masses of 4.4 kg for germanium (red) and 1.1 kg for silicon (blue). (a) Electric dipole with $d_\chi = 10^{-5}$ TeV$^{-1}$. (b) Magnetic dipole with $\mu_\chi = 10^{-3}$ TeV$^{-1}$. (c) Heavy broken $U(1)$ with $m_A = 10$ MeV and $\lambda = 10^{-7}$. (d) Light broken $U(1)$ with $m_A = 1$ meV and $\lambda = 10^{-14}$.}
\label{fig:rates}
\end{figure}

Results are shown in Figure \ref{fig:dipolelimits} for dipole moments and in Figure \ref{fig:u1limits} for broken $U(1)$ models, with sensitivities for silicon, germanium, and xenon detectors. As mentioned before, we consider the mass range $600$ keV $< m_\chi < 10$ GeV. The upper limit is due to current nuclear recoil experiments, most of which provide substantial exclusions down to $m_\chi \sim 10$ GeV \cite{BanksEtAl}. The lower limit is simply due to the limited energy available for recoils at such light masses, which suppresses the possible signal.

For each interaction, we show the strongest current experimental and astrophysical constraints on the relevant parameter space. There are also detailed constraints placed by supernova cooling \cite{FayetEtAl,ProfumoEtAl} and BBN \cite{BurlesEtAl} on DM with mass $m_\chi \lesssim 10$ MeV, but the full calculation of those constraints for these particular models is beyond the scope of this paper. For this reason, we simply indicate in our plots the mass value below which these additional bounds potentially apply. As a conceptual reference for the plots, the minimum $d_\chi$ and $\mu_\chi$ values excluded by germanium without background for $m_\chi = 10$ MeV correspond to $\langle \sigma v \rangle \approx 10^{-45}$ cm$^2$.

As discussed in the beginning of section \ref{Sec:Rate}, the larger binding energy present in xenon necessitates a larger momentum transfer. This results in an increased form factor suppression, reducing the experimental reach of noble gases in comparison with that of semiconductors, as seen in both Figures \ref{fig:dipolelimits} and \ref{fig:u1limits}.

For each of these exclusion limits, germanium provides somewhat weaker limits. This is caused by our assumption of the same background per unit mass for both materials, which places a stronger restriction on germanium, due to its heavier nuclear mass. Germanium's reach is also slightly reduced by its smaller effective electron mass in the conduction band, but does become more competitive with silicon at lower DM masses, due to its smaller band gap.

For EDM and MDM, the strongest general constraints on lighter masses come from colliders such as LEP \cite{FortinEtAl}. The parameter space for larger masses ($>1$ GeV) is also probed by the direct detection experiments XQC \cite{McCammonEtAl,SigurdsonEtAl}, CRESST \cite{AngloherEtAl}, CDMS \cite{AkeribEtAl}, and XENON10 \cite{AngleEtAl}. Direct detection rates for MDM are suppressed by the extra factor of $v^2$ relative to EDM limits. This effect applies to both current nuclear recoil experiments and CDMSLite, while collider limits do not suffer from this suppression \cite{SigurdsonErratum}.

Current dark matter annihilation searches such as Fermi LAT \cite{AtwoodEtAl}, and bounds on dark matter annihilation rates in the early universe from WMAP and ACT \cite{GalliEtAl} also provide limits to DM dipole moments. However, these results can only constrain symmetric DM, and do not apply to asymmetric models of DM. The resulting constraints are strong for MDM interactions, but the case of EDM annihilations is much more suppressed, making those irrelevant for our purposes.

As discussed in section \ref{Sec:Models}, models which generate dipole moments will also generate pointlike four-fermion interactions. Current bounds on these pointlike operators placed by collider experiments or current direct detection results could then potentially constrain dipole moments, as well. However, the charged particles which generate dipole moments will generically not couple directly to electrons or quarks. The generation of an effective vertex between DM and electrons would therefore require the exchange of a Z boson, in addition to the loop of charge intermediaries. The resulting effective operator would be substantially suppressed by a loop factor, the intermediary mass, and the Z boson propagator. Models which generate the dipole moments considered in this paper ($\Lambda \gtrsim 1$ TeV) would therefore be unconstrained by current bounds on pointlike four-fermion operators. However, the coupling of light DM to Z bosons which contributes to these operators is constrained by measurements of the Z width \cite{BeringerEtAl}. If the resulting Z dipole moments are approximately the same order as the EDM or MDM, we estimate the dipole moment scale would be constrained to be $\gtrsim 2$ TeV. These EDM and MDM bounds are very model-dependent, and therefore are not included in Figure \ref{fig:dipolelimits}. 

As mentioned earlier, there are two limiting cases for a broken $U(1)$. The first corresponds to a heavy mediator, for which we consider $m_A = 10$ MeV, and the second to a light mediator, for which we use $1$ meV. These sample masses were selected for comparison with \cite{EssigEtAl}.

The strongest constraints on $\lambda$, defined previously in equation (\ref{eqn:lambda}), come from a combination of astrophysical data, due to the fact that $\lambda$ depends on both the DM self-coupling and the $A'_\mu$-photon mixing. Details on the $U(1)$ mixing constraints can be found in \cite{JaeckelEtAl}, while the strongest constraints on DM self-coupling are explained in \cite{EssigEtAl,BuckleyEtAl,FengEtAl}. The larger mass regions are also limited by the same direct detection experiments as dipole moments.

In Figure \ref{fig:rates}, we also provide example detection rates for each model considered above, assuming a DM mass of 100 MeV, as well as the same runtime (1 year) and detector masses (4.4 kg for Ge, 1.1 kg for Si). In the case of larger background, annual modulation in these rates due to the Earth's motion around the Sun would provide an important confirmation of a potential DM signal. We estimate that the annual modulation would be approximately $6\%$ of the baseline average for the case of both EDM and MDM, which is larger than the traditional modulation associated with a velocity-independent cross-section. This difference arises due to the initial electron wavefunction, which provides substantial $q$-dependence (and therefore velocity-dependence) in the final cross-section.

\section{Conclusions}

We have examined the potential direct detection reach of semiconductor-based experiments such as CDMSLite. Experimental sensitivity to the production of single electron-hole pairs can dramatically improve the detectable energy range over that of traditional nuclear recoil methods. This enhanced energy range kinematically allows electron recoils to probe the parameter space of light dark matter, a region which remains largely inaccessible to nuclear recoils. The small energy gap present in semiconducting bandstructure also provides materials such as silicon and germanium with a substantial advantage relative to noble gases. Motivated by these prospects, we have considered the possible interactions between electrons and dark matter, many of which fit naturally into weak-scale extensions to the standard model.

We have found that semiconductor detectors are sensitive to a large range of uninvestigated parameter space, specifically for interactions such as dipole moments, which are enhanced at low recoil energies. Such dipole moments are naturally generated in many extensions of the standard model and are generically expected in models where the dark matter carries a conserved charge, such as asymmetric dark matter \cite{Nussinov,Kaplan,KitanoEtAl,KaplanEtAl,FalkowskiEtAl}. Electromagnetic moments provide a unique glimpse into higher-scale physics, and we have found that CDMSLite can extend our current reach by orders of magnitude, up to scales as large as $10^3$ TeV. Light dark matter with dipolar or new gauge interactions remains a well-motivated alternative to the traditional heavy WIMP scenario, and CDMSLite would present a substantial opportunity to explore this possibility.

\section*{Acknowledgments}
We would like to thank Blas Cabrera, Jeter Hall, Jeremy Mardon, and Matt Pyle for useful discussions. This work was supported in part by ERC grant BSMOXFORD no. 228169 and NSF grant PHY-0910467.

\section*{APPENDIX}

Here we present a more detailed description of ionization by dark matter. Our interaction system consists of the DM ($\vec{x}_\chi$), electron ($\vec{x}_e$), and nucleus ($\vec{x}_N$). Since the electron and nucleus originally form a bound state, we can change coordinate systems to more clearly show that this is a 2- to 3-body scattering process (bound atom + DM $\rightarrow$ electron + nucleus + DM). These coordinates are similar to the relative and center-of-mass coordinates used in strictly two-body systems. First we define new coordinates for the atomic electron-nucleus system,

\begin{equation}
\begin{split}
\vec{x}_a &= \vec{x}_e - \vec{x}_N \textrm{,} \\
\vec{x}_A &= \frac{m_e}{m_e+m_N} \vec{x}_e + \frac{m_N}{m_e+m_N} \vec{x}_N \textrm{.}
\end{split}
\end{equation}

We can then use these to define new coordinates for the full DM-atom system,

\begin{equation}
\begin{split}
\vec{x}_r &= \vec{x}_\chi - \vec{x}_A \textrm{,} \\
\vec{x}_R &= \frac{m_\chi}{m_\chi+m_A}\vec{x}_\chi + \frac{m_A}{m_\chi+m_A}\vec{x}_A \textrm{.}
\end{split}
\end{equation}

These new coordinates make it simple to write the system's initial state, as well as the relative velocity necessary for our scattering equation,

\begin{equation}
\begin{split}
\Psi_i(\vec{x}_a,\vec{x}_r,\vec{x}_R) &= \psi(\vec{x}_a) e^{i \vec{p}_r \cdot \vec{x}_r} e^{i \vec{p}_R \cdot \vec{x}_R} \textrm{,} \\
\vec{v}_{rel} = \frac{\vec{p}_r}{\mu_r} &= \frac{\vec{p}_\chi}{m_\chi} - \frac{\vec{p}_A}{m_A} \textrm{,}
\end{split}
\end{equation}

{\noindent}while the final state can easily be written in either coordinate system

\begin{equation}
\Psi_f(\vec{x}_a,\vec{x}_r,\vec{x}_R) = e^{i \vec{k}_a \cdot \vec{x}_a} e^{i \vec{k}_r \cdot \vec{x}_r} e^{i \vec{k}_R \cdot \vec{x}_R} = e^{i \vec{k}_e \cdot \vec{x}_e} e^{i \vec{k}_N \cdot \vec{x}_N} e^{i \vec{k}_\chi \cdot \vec{x}_\chi} \textrm{.}
\end{equation}

Now we should be more careful and worry about normalizing these free state wavefunctions, but whatever normalization factors we use now will cancel out and not affect our final interaction cross-section, so we will ignore them. We can now use these coordinates to simplify the matrix element for the electron recoil, assuming the interaction Hamiltonian $H$ is only a function of the relative displacement between the DM and electron

\begin{equation}
\left| \langle f|H|i \rangle \right|^2 = (2 \pi)^3 \delta^3 (\vec{k}_f - \vec{p}_i) \left| \tilde{\psi} \left(\vec{k}_a + \frac{\mu_a}{m_e}(\vec{k}_r-\vec{p}_r)\right) \right|^2 \, \left| \tilde{H}(\vec{k}_r-\vec{p}_r) \right|^2 \textrm{,}
\end{equation}

{\noindent}where $\tilde{H}(\vec{p})$ can be found by simply taking the nonrelativistic limit of the interaction amplitude $\mathcal{M}(\vec{p})$, with all spinors normalized to 1. The last assumptions we need to make to simplify our final cross-section formula is that the incoming momentum of the nucleus is zero and $m_e << m_N$. We can then integrate over the outgoing nuclear momentum and convert back to the more intuitive physical coordinates, resulting in

\begin{equation}
\begin{split}
d\sigma &= \frac{1}{|\vec{v}_{rel}|} \left( \displaystyle\prod_{f} \frac{d^3k_f}{(2 \pi)^3} \right) 2 \pi \, \delta(E_f - E_i) \left| \langle f|H|i \rangle \right|^2 \\
&\approx \frac{1}{|\vec{v}_\chi|} \, \frac{d^3 k_\chi}{(2\pi)^3} \, \frac{d^3 k_e}{(2\pi)^3} \, 2\pi \, \delta(E_f-E_i) \left| \tilde{\psi}(\vec{p}_\chi-\vec{k}_\chi-\vec{k}_e) \right|^2 \, \left| \tilde{H}(\vec{p}_\chi-\vec{k}_\chi) \right|^2 \textrm{.}
\end{split}
\end{equation}

In the limit $m_N \rightarrow \infty$, the fixed nucleus breaks translation invariance, and therefore momentum conservation. Perhaps a more straightforward interpretation is that the nucleus can absorb any finite momentum at negligible energy cost, due to the $m_N$ suppression in its kinetic energy. Either way, energy conservation is now the only constraint on the electron-DM system.

We can further simplify our final expression by taking the limit $k_e << q$, where $q$ is the magnitude of the momentum transfer ($\vec{q} = \vec{p}_\chi - \vec{k}_\chi$). We can check that this limit is valid by comparing the expressions for $k_e$ and the minimum possible $q$ for a given incoming $v_\chi$,

\begin{equation}
\begin{split}
k_e &= \sqrt{2 m_e E_R} \textrm{,} \\
q_{min} &= m_\chi v_\chi - \sqrt{m_\chi^2 v_\chi^2 - 2 m_\chi (E_R + E_B)} \textrm{.}
\end{split}
\end{equation}

We then see that $q_{min}$ can be further reduced either by increasing $m_\chi$ or $v_\chi$ or by decreasing $E_R$ or $E_B$. If we then consider the most extreme case, with $m_\chi = 10$ GeV, $v_\chi = v_{esc}$, $E_B = 1$ eV, and $E_R = 1$ eV, we find that $k_e \approx q_{min} \approx 1$ keV. However, the density of states for such a small binding energy is negligible, especially in the case of silicon. Increasing the binding energy to values with higher detection efficiency will only increase the ratio of $q_{min}$ to $k_e$, as will decreasing $m_\chi$ or $v_\chi$. For the bulk of the relevant parameter space, then, $k_e$ will be substantially smaller than even the minimum possible momentum transfer, making this approximation valid.

As a pedagogical example, we now choose a simple point-vertex interaction and use the ground-state hydrogen wavefunction for $\psi$. After enforcing energy conservation and integrating over the trivial angles, we find the expression

\begin{equation}
\frac{d\sigma}{dE_R} \approx \frac{16 a^3 k_e m_e k_\chi m_\chi}{\Lambda^4 \pi^2 v_\chi} \int dcos\theta_\chi (1+a^2(\vec{p}_\chi - \vec{k}_\chi)^2)^{-4} \textrm{.}
\end{equation}

The na\"{i}ve expectation would be for this integral to be $\mathcal{O}(1)$ at low energy recoil (and therefore low momentum transfer), but it turns out that there is a large suppression for most of the range of integration. This can be understood as a phase space suppression for any angle that deviates away from forward scattering (or alternatively, for any momentum exchange $q$ that deviates away from the minimum value $q_{min} = p_\chi - k_\chi$). This means that our final expressions will be dominated by terms corresponding to $q_{min}$. In this example, we obtain

\begin{equation}
\begin{split}
\frac{d\sigma}{dE_R} &\approx \frac{8 a k_e m_e}{3 \Lambda^4 \pi^2 v_\chi^2} [(1+a^2(p_\chi-k_\chi)^2)^{-3} - (1+a^2(p_\chi+k_\chi)^2)^{-3}] \\
&\approx \frac{8 a k_e m_e}{3 \Lambda^4 \pi^2 v_\chi^2} \frac{1}{(1+a^2(q_{min})^2)^3} \textrm{.}
\end{split}
\end{equation}

The second apparent form of suppression comes in the factor of $m_e$ in the numerator, rather than some form of atom-DM reduced mass. This can be understood by considering the limiting behavior as $m_e \rightarrow \infty$. In this limit, the cross-section rapidly grows for recoil energies $E_R \approx 0$, and is heavily suppressed for all others, converging on a delta function centered at zero recoil energy (an ionized electron with no kinetic energy). This limiting behavior should have been expected to appear, so it is not surprising to see the factor $k_e m_e$ in our final answer.

Applying this same approach to the models considered in this paper, we obtain the following full cross-sections

\begin{equation}
\begin{split}
\left. \frac{d \sigma}{dE_R} \, \right|_{EDM} \, = \, \frac{16 a^2 d_{\chi}^2 k _e}{\pi v_{\chi}^2} \, & \left[ \frac{6a^4(p_\chi+k_\chi)^4 + 15a^2(p_\chi+k_\chi)^2 + 11}{6(1+a^2(p_\chi+k_\chi)^2)^3} - 2 \ln \left( \frac{p_\chi-k_\chi}{p_\chi+k_\chi} \right) \right. \\
& \left. + \, \ln \left( \frac{1+a^2(p_\chi-k_\chi)^2}{1+a^2(p_\chi+k_\chi)^2} \right) - \frac{6a^4(p_\chi-k_\chi)^4 + 15a^2(p_\chi-k_\chi)^2 + 11}{6(1+a^2(p_\chi-k_\chi)^2)^3} \right] \textrm{,}
\end{split}
\end{equation}

\begin{equation}
\begin{split}
\left. \frac{d \sigma}{dE_R} \, \right|_{MDM} \, = \, \frac{64 \alpha^2 a^2 \mu_\chi^2 k_e}{3 \pi v_\chi^2} \, & \left[ \frac{6a^4(p_\chi+k_\chi)^4 + 15a^2(p_\chi+k_\chi)^2 + 11}{6(1+a^2(p_\chi+k_\chi)^2)^3} \right. - 2 \ln \left( \frac{p_\chi-k_\chi}{p_\chi+k_\chi} \right) \\
& + \, \ln \left( \frac{1+a^2(p_\chi-k_\chi)^2}{1+a^2(p_\chi+k_\chi)^2} \right) - \frac{6a^4(p_\chi-k_\chi)^4 + 15a^2(p_\chi-k_\chi)^2 + 11}{6(1+a^2(p_\chi-k_\chi)^2)^3} \\
& + \, \frac{1}{4} (1+a^2(p_\chi-k_\chi)^2)^{-3} - \left. \frac{1}{4}(1+a^2(p_\chi+k_\chi)^2)^{-3} \right] \textrm{,}
\end{split}
\end{equation}

\begin{equation}
\begin{split}
\left. \frac{d \sigma}{dE_R} \, \right|_{U(1)} \, = \, \frac{512 \lambda^2 a^4 k_e}{v_{\chi}^2 (a^2 m_A^2 - 1)^4} \, \left[ \frac{1}{(a^2 m_A^2 - 1)} \ln \left( \frac{(1+a^2(p_\chi-k_\chi)^2)(m_A^2+(p_\chi+k_\chi)^2)}{((1+a^2(p_\chi+k_\chi)^2)(m_A^2+ (p_\chi-k_\chi)^2)} \right) \right. \\
\left. + \, \frac{(a^2m_A^2-1)^2-3(a^2m_A^2-1)(1+a^2(p_\chi-k_\chi)^2)+9(1+a^2(p_\chi-k_\chi)^2)^2}{12(1+a^2(p_\chi-k_\chi)^2)^3} \right. \\
\left. - \, \frac{(a^2m_A^2-1)^2-3(a^2m_A^2-1)(1+a^2(p_\chi+k_\chi)^2)+9(1+a^2(p_\chi+k_\chi)^2)^2}{12(1+a^2(p_\chi+k_\chi)^2)^3} \right. \\
\left. + \, \frac{1}{4a^2(m_A^2 + (p_\chi-k_\chi)^2)} - \frac{1}{4a^2(m_A^2 + (p_\chi+k_\chi)^2)} \right] \textrm{.}
\end{split}
\end{equation}

If we then take the appropriate limits of these cross-sections, we obtain the approximate results given earlier.

\bibliography{References}
\bibliographystyle{elsarticle-num}

\end{document}